\documentclass[final,3p,times,twocolumn]{elsarticle}

\usepackage{amssymb}

\usepackage{amsfonts}
\usepackage{amsmath}
\usepackage{xcolor,tikz,pgfplots}
\usepackage{hyperref} 

\usepackage{float}
\usepackage{booktabs}
\usepackage{caption}

\journal{Physics Letters B}

\begin{document}

\begin{frontmatter}


\title{Infrared Duality in Unoriented Pseudo del Pezzo}

\author[label1]{Andrea Antinucci\fnref{1}}
\author[label1,label2]{Salvo Mancani\fnref{2}}
\author[label2]{Fabio Riccioni\fnref{3}}

\address[label1]{Dipartimento di Fisica, Universit\`a di Roma “La Sapienza”, Piazzale Aldo Moro 2, 00185 Roma, Italy}
\address[label2]{INFN Sezione Roma1, Dipartimento di Fisica, Universit\`a di Roma “La Sapienza”, Piazzale Aldo Moro 2, 00185 Roma, Italy}

\fntext[1]{antinucci.1711393@studenti.uniroma1.it}
\fntext[2]{salvo.mancani@uniroma1.it}
\fntext[3]{fabio.riccioni@roma1.infn.it}

\begin{abstract}
We study the orientifold projections of the $\mathcal{N}=1$ superconformal field theories describing D3-branes probing the Pseudo del Pezzo singularities PdP$_{3b}$ and PdP$_{3c}$. The PdP$_{3c}$ parent theory admits two inequivalent orientifolds. Exploiting $a$ maximization, we find that one of the two has an $a$-charge smaller than what one would expect from the orientifold projection, which suggests that the theory flows to the fixed point in the infrared. Surprisingly, the value of $a$ coincides with the charge of the unoriented PdP$_{3b}$ and we interpret this as the sign of an infrared duality.
\end{abstract}


\end{frontmatter}

\section{Introduction}

The AdS/CFT correspondence, in its original formulation, states that the 4 dimensional $\mathcal{N}=4$ $\textrm{SU}(N)$ SYM gauge theory living on the worldvolume of a stack of $N$ D3-branes in flat space is dual to type-IIB supergravity on an AdS$_5 \times S^5$ background in the large $N$ limit~\cite{Maldacena_1999, Gubser_1998, Witten:1998qj}. More generally, for a system of regular D3-branes probing the tip of a Calabi-Yau (CY) cone, the worldvolume conformal field theory is dual to IIB supergravity on AdS$_5 \times H^5$, where the horizon $H^5$ is a 5-dimensional Sasaki-Einstein manifold~\cite{Morrison:1998cs, Klebanov:1998hh} and represents the base of the CY cone. The correspondence relates the central charge $a$ of the conformal field theory 
\begin{equation}\label{eq:a-charge}
a = \frac{3}{32} \left( 3 \mathrm{Tr}R^3 - \mathrm{Tr}{R} \right) \;  \ , 
\end{equation}
to the volume of the Sasaki-Einstein horizon $H^5$ by the relation~\cite{Gubser_1998amax}
\begin{equation}\label{eq:Volume}
\mathrm{Vol}(H^5) = \frac{\pi^3}{4} \frac{N^2}{a} \; ,
\end{equation}
where $N$ is the number of units of 5-form flux. In general, the presence of $\textrm{U}(1)$ flavour symmetries implies that the $R$-charges can not be unambiguously assigned a priori.  When this happens, the superconformal $R$-charges are uniquely determined as those that maximize $a$~\cite{Intriligator:2003jj, Bertolini:2004xf}. 

\begin{figure}[h]
\centering{\includegraphics[scale=0.35, trim={4cm 4.3cm 5cm 2.5cm}, clip]{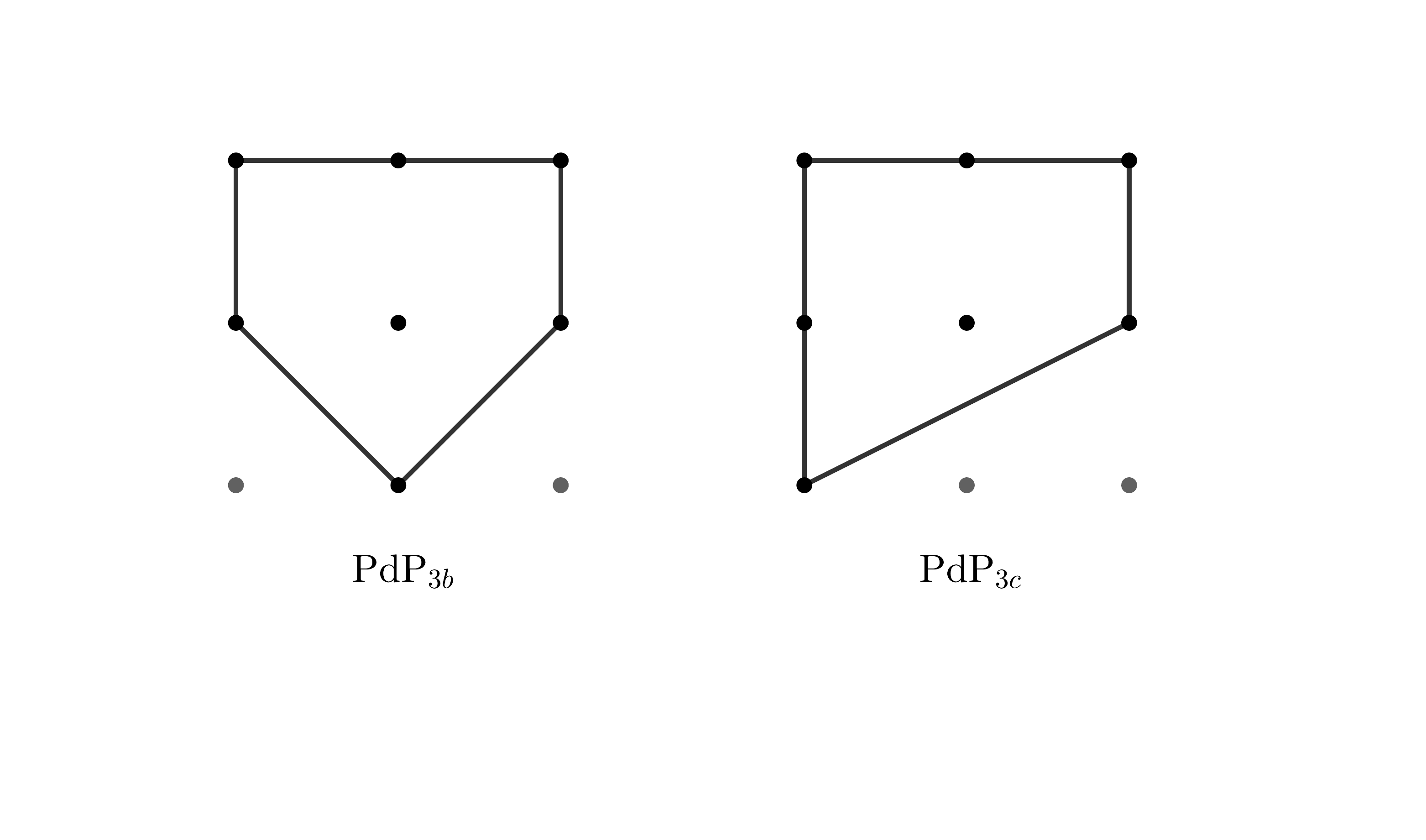}}
\caption{The toric diagram of PdP$_{3b}$ on the left and the toric diagram of PdP$_{3c}$ on the right.}\label{fig:PdP3bcToric}
\end{figure}

While in the simpler cases of cones that are abelian orbifolds of $\mathbb{C}^3$ the field content and superpotential of the gauge theory can be read directly from the $\mathcal{N}=4$ ones, systematic techniques have been developed to determine the gauge theories for D3-branes at the tip of more general \emph{toric} cones~\cite{Beasley_2000}. A toric cone has at least a $\textrm{U}(1)^3$ isometry and its base $H^5$ is a $\mathbb{T}^3$ fibration over a convex polygon known as the \emph{toric diagram}. In particular, one can study the blow-up of a singularity adding points to the toric diagram and this is dual to the unhiggsing mechanism in the gauge theory~\cite{Morrison:1998cs, Beasley_2000, Uranga:1998vf, Feng_2003, Feng_2004}. If applied to the dP$_2$ theory, corresponding to the complex cone over the del Pezzo surface obtained by the blow-up of 2 generic points of $\mathbb{P}^2$, this gives rise to either dP$_3$ or the Pseudo del Pezzo geometries PdP$_{3b}$ or PdP$_{3c}$ if the blow-up is non-generic~\cite{Feng_2003, Feng_2004}. The latter two theories, whose toric diagrams are drawn in Fig.~\ref{fig:PdP3bcToric}, are the ones we are interested in. 

From the toric diagrams, there is a systematic procedure to determine, up to Seiberg dualities, the field content and the superpotential of the gauge theory. Specifically, this information is encoded in a bipartite graph representing a web of NS5-branes, which is called  \emph{dimer}~\cite{Franco:2005sm, Franco_2006, hanany2005dimer, Franco:2012mm, Franco:2013ana}. This web is drawn on a $\mathbb{T}^2$ wrapped by D5-branes, and by performing two T-dualities along the torus the whole configuration is mapped back to the original system of D3-branes sitting at the singularity of the cone. We draw in Figs.~\ref{fig:PdP3bDimer} and~\ref{fig:PdP3cDimer} the dimers of the PdP$_{3b}$ and PdP$_{3c}$ theories. By looking at the nodes in the diagrams one can see that the two theories have different superpotentials but share the same quiver, given in Fig.~\ref{fig:quiverPdP3bc}. 

In this setup, more general gauge theories can be constructed by introducing orientifold projections $\Omega$. In the brane configuration, this corresponds to adding an \emph{orientifold plane}, which induces a $\mathbb{Z}_2$ involution on the space and flips the world-sheet parity of strings, making the theory unoriented~\cite{Sagnotti:1987tw, Pradisi:1988xd, Bianchi:1990yu, Bianchi:1990tb, Polchinski:1995mt} (for a review see~\cite{Angelantonj:2002ct}). The $\mathbb{Z}_2$ involution of the orientifold can be represented on the dimer by the introduction of either fixed points or fixed lines~\cite{Franco:2007ii}, from which one can read the field content and the superpotential of the unoriented theory. As far as conformal invariance is concerned, the following scenarios are possible: either the unoriented theory does not have a fixed point at which $a$ is maximized, or there is a fixed point and the orientifold yields $O(1/N)$ corrections to physical observables. Note that in the first scenario one could have a duality cascade~\cite{Argurio:2017upa} or conformal symmetry can be restored by the addition of flavour branes~\cite{Bianchi:2013gka}.

In this letter we show that a third scenario exists, in which the orientifold breaks the conformal symmetry of the parent theory but develops a new superconformal fixed point. We provide an example of such a scenario by studying the orientifold projection of the gauge theories arising from the toric CY cones over PdP$_{3b}$ and PdP$_{3c}$. While the projection PdP$_{3b}^{\Omega}$ is unique, PdP$_{3c}$ admits two inequivalent orientifolds, which we call $\Omega_1$ and $\Omega_2$. For each of the orientifolds, we determine the ranks of the gauge groups such that there is a conformal fixed point, and the superconformal $R$-charges of the chiral fields using $a$ maximization. Although one would always naively expect that the $a$ central charge of the unoriented theory is a half of that of the parent theory, we find that this happens only for PdP$_{3b}^{\Omega}$ and PdP$_{3c}^{\Omega_1}$, while for PdP$_{3c}^{\Omega_2}$ one gets a smaller central charge $a$. To the best of our knowledge, this is the first time that such a mechanism occurs in the context of unoriented theories. Interestingly, both the $R$-charges and the $a$ central charge of PdP$_{3c}^{\Omega_2}$ are identical to the ones of PdP$_{3b}^{\Omega}$. To see this effect it is crucial that the computation is performed keeping $N$ finite, while neglecting $O(1/N)$ corrections naively gives the same $a$ central charge of PdP$_{3c}^{\Omega_1}$. The fact that the central charge $a$ is smaller than expected suggests that there is an RG flow from a conformal fixed point in the UV to another conformal fixed point in the IR, in which the theory coincides with PdP$_{3b}^{\Omega}$. This IR duality could be inherited by the relation between the parent theories, which are connected by a web of dualities involving specular~\cite{Hanany:2012hi, Hanany_2012} and Seiberg duality~\cite{Seiberg:1994pq, Beasley:2001zp}.

\section{PdP$_{3b}$ and PdP$_{3c}$}

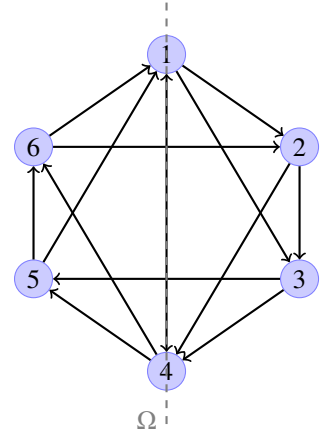
\begin{figure}
\centering
{\begin{tikzpicture}[auto, scale= 0.35]
        \node [circle, draw=blue!50, fill=blue!20, inner sep=0pt, minimum size=5mm] (3) at (5,-2.5) {$3$};            
        \node [circle, draw=blue!50, fill=blue!20, inner sep=0pt, minimum size=5mm] (5) at (-5,-2.5) {$5$};
        \node [circle, draw=blue!50, fill=blue!20, inner sep=0pt, minimum size=5mm] (6) at (-5,2.5) {$6$}; 
        \node [circle, draw=blue!50, fill=blue!20, inner sep=0pt, minimum size=5mm] (2) at (5,2.5) {$2$}; 
        \node [circle, draw=blue!50, fill=blue!20, inner sep=0pt, minimum size=5mm] (1) at (0,6) {$1$};
        \node [circle, draw=blue!50, fill=blue!20, inner sep=0pt, minimum size=5mm] (4) at (0,-6) {$4$}; 
        \draw (1)  to node {} (2) [->, thick, ];
        \draw (2)  to node {} (3) [->, thick, ];
        \draw (3)  to node {} (4) [->, thick, ];     
        \draw (4)  to node {} (5) [->, thick, ];
        \draw (5)  to node {} (6) [->, thick, ];
        \draw (6)  to node {} (1) [->, thick, ];
        \draw (1)  to node {} (4) [<->, thick, ];
        \draw (6)  to node {} (2) [->, thick, ];
        \draw (2)  to node {} (4) [->, thick, ];
        \draw (4)  to node {} (6) [->, thick, ];
        \draw (3)  to node {} (5) [->, thick, ];
        \draw (5)  to node {} (1) [->, thick, ];
        \draw (1)  to node {} (3) [->, thick, ];

        \draw [thick, dashed, gray] (0, -8) to node [pos=0.01] {$\Omega$} (0,8) ;
        \end{tikzpicture}}
        \caption{The quiver of theories PdP$_{3b}$ and PdP$_{3c}$. The dashed gray line labelled as $\Omega$ represents the orientifold projection, which identifies the two sides of the quiver and projects fields and gauge groups that lie on top of it.} \label{fig:quiverPdP3bc}
\end{figure}

\begin{figure}[b]
\centering
\includegraphics[scale=0.3, trim={4cm 2cm 9.4cm 2.1cm}, clip]{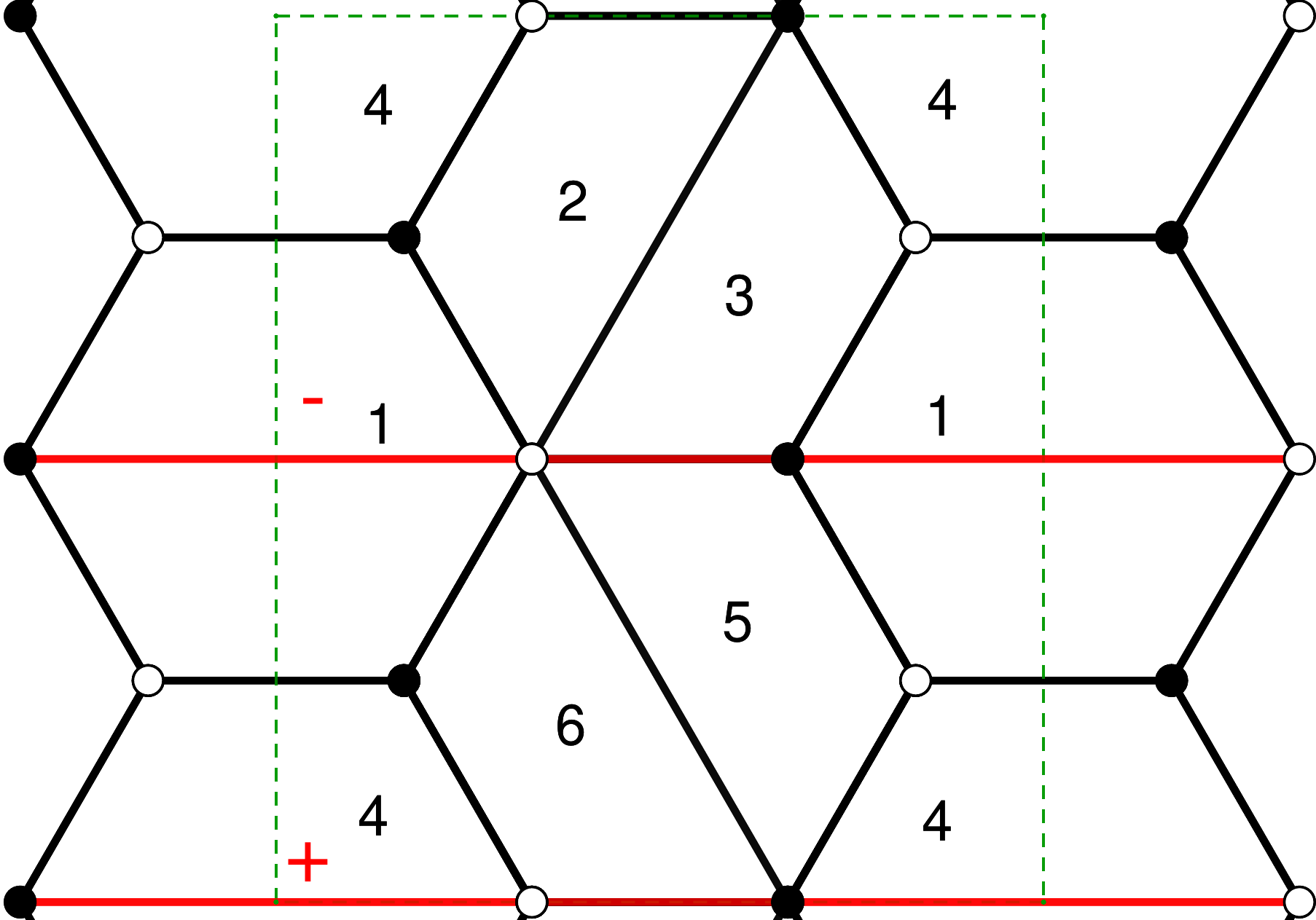}
\caption{The dimer of PdP$_{3b}$, where the dashed green line delimits the fundamental cell. The two red fixed lines and their signs represent the orientifold projection that yields the unoriented PdP$_{3b}^{\Omega}$.}\label{fig:PdP3bDimer}
\end{figure}

Let us introduce the parent gauge theories of interest. We begin with PdP$_{3b}$~\cite{Feng_2003, Feng_2004, Hanany:2012hi}, whose quiver and dimer are drawn in Fig.~\ref{fig:quiverPdP3bc} and~\ref{fig:PdP3bDimer}. There are 6 gauge groups $ \prod_{a=1}^6 \textrm{SU}(N_a) $ and the matter fields are bifundamentals $X_{ab}$ corresponding to the edges in the dimer. For instance, $X_{12}$ transforms in the fundamental representation of $\textrm{SU}(N_1)$ and in the anti-fundamental of $\textrm{SU}(N_2)$ and is the edge between faces 1 and 2 in the dimer. The global symmetries of this model are $\textrm{U}(1)^2 \times \textrm{U}(1)_R$ as mesonic symmetry and $\textrm{U}(1)^5$ as baryonic one, of which $\textrm{U}(1)^2$ is anomalous. The superpotential of the theory reads
\begin{align}
W_{3b} &= X_{13} X_{34} X_{41} - X_{46} X_{61} X_{14} + X_{45} X_{51} X_{14} \nonumber \\
& - X_{24} X_{41} X_{12} + X_{62} X_{24} X_{46} - X_{35} X_{51} X_{13} \nonumber \\
& + X_{23} X_{35} X_{56}X_{61} X_{12} - X_{23} X_{34} X_{45} X_{56} X_{62} \; .
\end{align}
The gauge anomalies vanish imposing the following relation between the ranks of the gauge groups:
\begin{align}\label{eq:GaugeAnomalies}
N_1 + N_6 - N_3 - N_4 &= 0 \; , \nonumber \\
N_2 + N_3 - N_5 - N_6 &= 0\; .
\end{align}
We find the superconformal fixed point and the corresponding $R$-charges $R_{ab}$ for the fields $X_{ab}$ maximizing the $a$-charge. Requiring that the $\beta$-functions vanish (which is equivalent to non-anomalous $R$-symmetry) we have
\begin{equation}
\sum_a (R_{ab} - 1)N_a = -2 N_b \; ,
\end{equation}
where the sum is over gauge groups $a$ connected to $b$ by a bifundamental field $X_{ab}$. Together with the condition that the $R$-charge of the superpotential is $R(W)=2$, we have a system of equations with a priori eight independent $R$-charges. This can be seen also from the quiver, which enjoys a $\mathbb{Z}_2$ symmetry. The $a$-charge in Eq.~\eqref{eq:a-charge} is a two-variable function, namely, a flavour symmetry $\textrm{U}(1)^2$ mixes with the $R$-symmetry. The local maximum yields~\cite{Hanany:2012hi}
\begin{align}\label{eq:RPdP3b}
R_{23}^{b} &= 7 - 3 \sqrt{5} \; , \nonumber \\
R_{13}^{b} &= R_{14}^{b} = R_{24}^{b} = 3 - \sqrt{5} \; , \nonumber \\ 
R_{12}^{b} &= R_{34}^{b} = R_{35}^{b} = R_{62}^{b} = 2 \sqrt{5} - 4 \; , 
\end{align}
\begin{align}
&a_{3b} = \frac{27}{4} N^2 \left( 5 \sqrt{5} - 11 \right) \; , \nonumber \\
&\mathrm{Vol}(\textrm{PdP}_{3b}) = \frac{\pi^3}{27 \left( 5 \sqrt{5} - 11 \right)} \; ,
\end{align}
where $N_a = N$ $\forall a = 1, \ldots , 6$ since this condition gives the only solution that respects the unitarity bound. Note that the expression of the $a$-charge is given at leading order in $N$. 

\begin{figure}
\centering
\includegraphics[scale=0.35, trim={14.5cm 2.85cm 16.2cm 2.95cm}, clip]{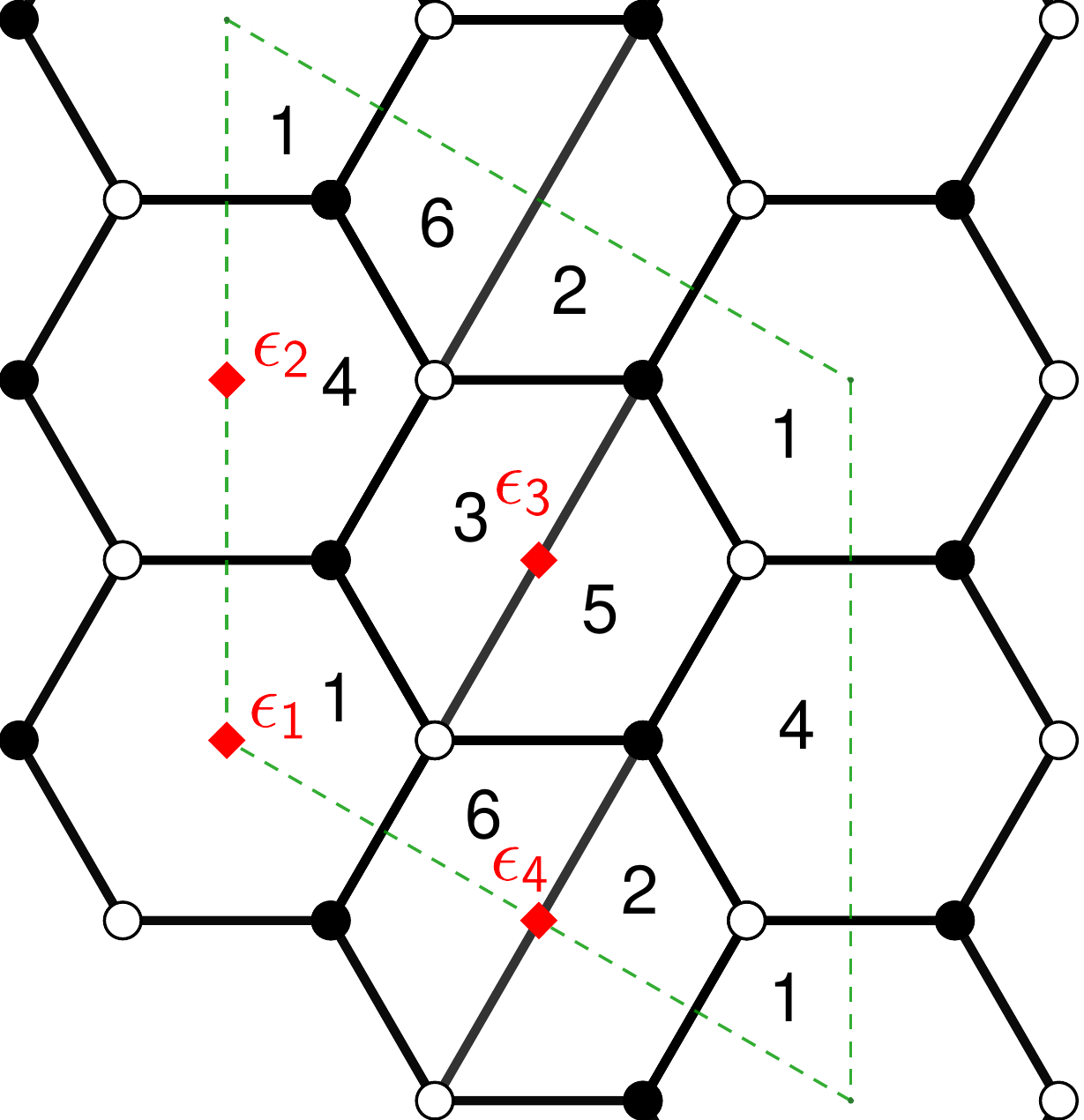}
\caption{The dimer of PdP$_{3c}$, where the dashed green line delimits the fundamental cell. The four red fixed points $(\epsilon_1, \epsilon_2, \epsilon_3, \epsilon_4)$ represent the orientifold projection, where $(+,-,-,+)$ corresponds to PdP$_{3c}^{\Omega_1}$ and $(-,+,-,+)$ corresponds to PdP$_{3c}^{\Omega_2}$.}\label{fig:PdP3cDimer}
\end{figure}

The second theory we study is PdP$_{3c}$~\cite{Feng_2003, Feng_2004, Hanany:2012hi}, whose dimer is drawn in Fig.~\ref{fig:PdP3cDimer}. As in the previous case, the gauge group is $\prod_{a=1}^6 \textrm{SU}(N_a)$ and the global symmetries are $\textrm{U}(1)^2 \times \textrm{U}(1)_R$ as mesonic symmetry and $\textrm{U}(1)^5$ as baryonic one, of which $\textrm{U}(1)^2$ is anomalous. The matter fields are also the same of the PdP$_{3b}$ theory and indeed the two models share the same quiver in Fig.~\ref{fig:quiverPdP3bc}.  Nonetheless, they have a different dimer and therefore  interact differently. In fact, the superpotential reads
\begin{align}
W_{3c} &= X_{12} X_{24} X_{41} + X_{45} X_{51} X_{14} - X_{13} X_{34} X_{41} \nonumber \\
& - X_{46} X_{61} X_{14} + X_{13} X_{35} X_{56} X_{61} + X_{46} X_{62} X_{23} X_{34} \nonumber \\
& - X_{12} X_{23} X_{35}X_{51} - X_{45} X_{56} X_{62} X_{24} \; .
\end{align}
The gauge anomalies vanish imposing the condition in Eq.~\eqref{eq:GaugeAnomalies} as before. Computing the $R$-charges which maximize the $a$-charge we find~\cite{Hanany:2012hi}
\begin{align}\label{eq:RPdP3c}
R_{14}^{c} &= 2 - \frac{2\sqrt{3}}{3} \; , \nonumber \\
R_{23}^{c} &= R_{35}^{c} = R_{62}^{c} = 1 - \frac{\sqrt{3}}{3} \; , \nonumber \\
R_{12}^{c} &= R_{13}^{c} = R_{24}^{c} = R_{34}^{c} = \frac{\sqrt{3}}{3} \; , 
\end{align}
\begin{align}
&a_{3c} = \frac{3 \sqrt{3}}{4} N^2  \; , \nonumber \\
&\mathrm{Vol}(\textrm{PdP}_{3c}) = \frac{\pi^3}{3 \sqrt{3}} \; ,
\end{align}
where again $N_a = N$ $\forall a = 1, \ldots , 6$ is the only solution that respects the unitarity bound. Note that the difference in the $R$-charges arises from the condition $R(W)=2$. It is crucial to note that in this case the $a$-charge is a three-variable function, namely, the non-$R$ symmetry which mixes with the $R$-charge is $\textrm{U}(1)^3$. 

\section{Unoriented PdP$_{3b}$ and PdP$_{3c}$}

The orientifold projection of a quiver gauge theory is represented as a line which identifies the two sides  of the quiver  and projects the groups and the fields that are mapped onto themselves~\cite{Bianchi:2007wy, Bianchi:2009bg, Bianchi:2013gka}. For the PdP$_{3b}$ and PdP$_{3c}$ theories, the involution is the $\Omega$ line  in Fig.~\ref{fig:quiverPdP3bc}. The gauge groups intersected by the line are projected onto either SO or Sp, while bifundamental fields charged under two groups that are identified by $\Omega$ are projected onto a symmetric or antisymmetric representation. Besides, if $\textrm{SU}(N_a)$ is mapped to $\textrm{SU}(N_b)$, a fundamental representation of $\textrm{SU}(N_a)$ is identified with the antifundamental of $\textrm{SU}(N_b)$.

In the dimer, the  $\mathbb{Z}_2$ involution can be realized by the introduction of either fixed points of fixed lines on the torus~\cite{Franco:2007ii}. In the first case, the orientifold projects both coordinates of the torus, preserving the mesonic flavour symmetry.
There are four fixed points in the fundamental cell and each one carries a sign $\epsilon_i = \pm$, $i=1,\ldots,4$. The superpotential terms are halved since black nodes are mapped to white nodes. The four fixed points are physically four orientifold planes and their charges are the signs of the fixed points. When a fixed point with $\epsilon = + (-)$ lies on a face, the corresponding $\textrm{SU}(N)$ gauge group is projected onto $\textrm{SO}(\textrm{Sp})$. When a fixed point with $ \epsilon = + (-)$ lies on an edge it projects the corresponding field onto a symmetric (antisymmetric) representation and it identifies the gauge groups on the two sides of the edge. Supersymmetry requires that $\prod_{i=1}^4 \epsilon_i = (-1)^{N_W / 2}$, where $N_W$ is the number of terms in the superpotential of the parent theory. Besides, the anomaly cancellation condition may further constrain the orientifold signs.
In the second case, the involution acts projecting only one coordinate of the torus, thus breaking the $\textrm{U}(1)^2$ flavour symmetry to a combination of the two. Nodes in the dimer are mapped to each other w.r.t. the fixed lines. Physically, a fixed line is an orientifold plane that intersects the torus and its charge is the sign of the line. There is no constraint in this case, other than the anomaly cancellation condition.

\subsection{Unoriented PdP$_{3b}$} 

We consider the orientifold involution of PdP$_{3b}$ with two fixed lines and we choose the configuration with signs $(-,+)$, as in Fig.~\ref{fig:PdP3bDimer}. As a consequence, the gauge group $\textrm{SU}(N_5)$ is identified with $\textrm{SU}(N_3)$ and  $\textrm{SU}(N_6)$ with $\textrm{SU}(N_2)$, while $\textrm{SU}(N_1)$ becomes $\textrm{Sp}(N_1)$ and $\textrm{SU}(N_4)$ becomes $\textrm{SO}(N_4)$ since they lie on top of the fixed lines. The resulting theory has gauge groups $\textrm{Sp}(N_1) \times \textrm{SU}(N_2) \times \textrm{SU}(N_3) \times \textrm{SO}(N_4)$, where the fields $X_{35}^A$ and $X_{62}^S$ belong to the antisymmetric and symmetric representations of the gauge groups $\textrm{SU}(N_3)$ and $\textrm{SU}(N_2)$ respectively. The superpotential reads
\begin{align}
W_{3b}^{\Omega} &= X_{13} X_{34} X_{41} - X_{24} X_{41} X_{12} + X_{62}^S X_{24} X_{46} \nonumber \\ 
&- X_{35}^A X_{51} X_{13} + X_{23} X_{35}^A X_{56}X_{61} X_{12} \nonumber \\
&- X_{23} X_{34} X_{45} X_{56} X_{62}^S \;
\end{align}
and the anomaly cancellation condition is
\begin{align}\label{eq:AnomalyOmega}
N_1 + N_2 - N_3 - N_4 + 4 = 0 \; .
\end{align}
As in the parent theory, there are eight a priori independent $R_{ab}$. The condition for the $\beta$-functions to vanish changes slightly due to the orientifold involution. In fact, one has
\begin{equation}
\sum_a (R_{ab} - 1)N_a = -(N_b \pm 2) \; ,
\end{equation}
if the group labelled by $b$ is orthogonal $(-)$ or symplectic $(+)$. Likewise, for unitary gauge groups one has
\begin{equation}
\sum_a (R_{ab} - 1)N_a + (R_{b} - 1) (N_b \pm 2) = -2 N_b \; ,
\end{equation}
where $R_b$ is the $R$-charge of the symmetric $(+)$ or the antisymmetric $(-)$ field charged under $\textrm{SU}(N_b)$. Imposing conformal invariance, consistency with the unitarity bound requires $N_2 = N_3 = N_1 + 2 = N_4 - 2 = N$. One then finds the flavour symmetry is $\textrm{U}(1)^2$ and the $R$-charges are the same as the ones of the parent theory in Eq.~\eqref{eq:RPdP3b} up to $O(1/N)$ corrections. This implies that this orientifold realizes what we have described  in the introduction as the `second scenario', in which there is a fixed point, and the orientifold induces corrections to the $R$-charges that vanish at large $N$.
Taking this limit, the value of the $a$-charge is
\begin{align}\label{eq:aPdP3b}
a^{\Omega}_{3b} = \frac{27}{8} N^2 \left( 5\sqrt{5} - 11 \right) \; .
\end{align}
Note that the central charge $a$ is half the one of the parent theory. This is expected since the degrees of freedom have been halved. Besides, also the volume of the PdP$_{3b}$ after the orientifold is half the volume of the parent space. To see this, consider that the orientifold acts as a $\mathbb{Z}_2$ projection on the geometry. As a consequence, the number of units of 5-form flux is $N/2$~\cite{Witten:1998xy}. Thus, the proper ratio between the volumes reads
\begin{align}
\frac{\mathrm{Vol}(\textrm{PdP}^{\Omega}_{3b})}{\mathrm{Vol}(\textrm{PdP}_{3b})} = \frac{1}{2} \; .
\end{align}
This is similar to the case of the $\mathbb{Z}_n$ orbifold of flat space, where the volume is a fraction $n$ of the volume of the sphere $S^5$.

\subsection{Unoriented PdP$_{3c}$} 

Let us turn to the orientifold of PdP$_{3c}$. As shown in Fig.~\ref{fig:PdP3cDimer}, the dimer admits only the projection with fixed points, whose signs $(\epsilon_1,\epsilon_2,\epsilon_3,\epsilon_4)$ project the group $\textrm{SU}(N_1)$, the group $\textrm{SU}(N_4)$, the field $X_{35}$ and the field $X_{62}$, respectively. The parent theory has $N_W = 8$, thus $\prod_{i=1}^4 \epsilon_i = + 1$. The two inequivalent choices are $\Omega_1 = (+,-,-,+)$ and $\Omega_2 = (-,+,-,+)$. 

First, we focus on $\Omega_1$. The unoriented theory has gauge groups $\textrm{SO}(N_1) \times \textrm{SU}(N_2) \times \textrm{SU}(N_3) \times \textrm{Sp}(N_4)$, where fields $X_{35}^A$ and $X_{62}^S$ are antisymmetric and symmetric representations of $\textrm{SU}(N_3)$ and $\textrm{SU}(N_2)$ respectively. The superpotential reads
\begin{align}\label{eq:W3cOmega}
W_{3c}^{\Omega_1} &= X_{13} X_{35}^A X_{56} X_{61} - X_{45} X_{56} X_{62}^S X_{24} \nonumber \\ 
& + X_{12} X_{24} X_{41} - X_{13} X_{34} X_{41} \; 
\end{align}
and the anomaly cancellation condition remains as Eq.~\eqref{eq:AnomalyOmega}. The superconformal fixed point of this unoriented model has the same $R$-charges of the parent theory in Eq.~\eqref{eq:RPdP3c} up to $O(1/N)$ corrections, with the flavour $\textrm{U}(1)^3$ symmetry mixing with $R$-symmetry, and thus the second scenario described in the introduction is again realized as in the PdP$_{3b}$  case. In the large $N$ limit, the $a$-charge is
\begin{equation}\label{eq:Omega1a}
a_{3c}^{\Omega_1} = \frac{3 \sqrt{3}}{8} N^2 \; ,
\end{equation} 
where $N_1 = N_2 = N_3 - 2 = N_4 - 2 = N$ is the only consistent solution. Again, the $a$-charge is halved because of the orientifold projection, and the ratio between the volumes is $1/2$ as before.

The unoriented theory obtained from $\Omega_2 = (-,+,-,+)$ has gauge groups $\textrm{Sp}(N_1) \times \textrm{SU}(N_2) \times \textrm{SU}(N_3) \times \textrm{SO}(N_4)$, where fields $X_{35}^A$ and $X_{62}^S$ are unchanged w.r.t. the previous case. The superpotential $W_{3c}^{\Omega_2}$ is formally identical to $W_{3c}^{\Omega_1}$ in Eq.~\eqref{eq:W3cOmega} and again the anomaly cancellation condition remains as Eq.~\eqref{eq:AnomalyOmega}. The $a$-maximization in this case is more subtle. If one took naively the limit $N \to \infty$ before solving the equation for vanishing $\beta$-functions and $R(W)=2$, one would obtain the $R$-charges and half the $a$-charge of the parent theory, with non-$R$ flavour $\textrm{U}(1)^3$. On the other hand, we find that for any \emph{finite} value of $N$, the only consistent solution is $N_2 = N_3 = N_1 + 2 = N_4 - 2 = N$ exactly as in PdP$_{3b}^{\Omega}$, with one flavour $\textrm{U}(1)$ broken and the remaining $\textrm{U}(1)^2$ mixing with $R$-symmetry. This has the crucial effect of giving at leading order in $1/N$ the value of the superconformal $R$-charges as 
\begin{align}\label{eq:Omega2R}
R_{23}^{\Omega_2} &= 7 - 3\sqrt{5} \; , \nonumber \\
R_{13}^{\Omega_2} &= R_{14}^{\Omega_2} = R_{24}^{\Omega_2} = 3 - \sqrt{5} \; , \nonumber \\ 
R_{12}^{\Omega_2} &= R_{34}^{\Omega_2} = R_{35}^{\Omega_2} = R_{62}^{\Omega_2} = 2 \sqrt{5} - 4 \;  ,
\end{align}
which are different from the $R$-charges of the parent theory in Eq.~\eqref{eq:RPdP3c}, and the $a_{3c}^{\Omega_2}$-charge takes the value
\begin{align}\label{eq:Omega2a}
a_{3c}^{\Omega_2} &= \frac{27}{8} N^2 \left( 5 \sqrt{5} - 11 \right)  \; ,
\end{align}
which is smaller than the value of $a_{3c}^{\Omega_1}$ in Eq.~\eqref{eq:Omega1a}. Consequently, the ratio between $a_{3c}^{\Omega_2}$ and that of the parent theory is
\begin{equation}\label{eq:RatioOmega2a}
\frac{a_{3c}^{\Omega_2}}{a_{3c}} = \frac{3\sqrt{3}}{2} \left( 5 \sqrt{5} - 11 \right) \simeq 0.47 \; ,
\end{equation}
while the ratio between the volumes is
\begin{equation}
\frac{\mathrm{Vol}(\mathrm{PdP}_{3c}^{\Omega_2})}{\mathrm{Vol}(\mathrm{PdP}_{3c})} \simeq 0.53 \; .
\end{equation}

The fact that $a_{3c}^{\Omega_2}$ is less than halved w.r.t. the central charge $a_{3c}$ of the parent theory can be taken as a sign of an RG flow towards the IR. Thus, a natural question is what is the endpoint of this RG flow. Surprisingly, the $R$-charges and the $a$-charge in Eqs.~\eqref{eq:Omega2R} and~\eqref{eq:Omega2a} are exactly those of PdP$_{3b}^{\Omega}$ given in Eqs.~\eqref{eq:RPdP3b} and~\eqref{eq:aPdP3b}. This suggests the two theories are dual at the conformal fixed point. In other words, the RG flow is going from PdP$_{3c}^{\Omega_2}$ in the UV to PdP$_{3b}^{\Omega}$ in the IR. 

To further support this conjecture we investigate the $1/N$ corrections to the $R$-charges. Remarkably, imposing that the $\beta$-functions vanish yields exactly the same solutions at any finite $N$, which implies that the charges w.r.t. all the global symmetries of the two theories are the same. The values of these charges are reported in Tab.~\ref{tab}, where $Q_1$ and $Q_2$ are associated to the flavour symmetry $U(1)_1 \times U(1)_2$, while $R_0$ is an allowed non-superconformal choice of the $R$-charge.

\begin{center}
\vspace{15pt}
\begin{tabular*}{0.4\textwidth}{@{\extracolsep{\fill}}cccc}
\toprule
 & $Q_1$ & $Q_2$ &  $R_0$\\
\midrule
 $X_{12}$ & $-\frac{N+2}{N}$ &$\frac{1}{2}$ &   $\frac{1}{2}$  \\[3pt]
 $X_{13}$ & $0$ & $-\frac{1}{2}$ &   $\frac{1}{2}$  \\[3pt]
 $X_{24}$ & $\frac{1}{N}$ & $-\frac{1}{2}$  & $\frac{1}{2}$ \\[3pt]
 $X_{62}^S$ & $-\frac{4}{N}$ & $1$ & $1$\\[3pt]
 $X_{23}$ & $\frac{N+2}{N}$ & $-1$ & $0$ \\[3pt]
 $X_{34}$ & $-1$ & $\frac{1}{2}$ & $\frac{1}{2}$ \\[3pt]
 $X_{41}$ & $1$ & $0$ & $1$\\[3pt]
 $X_{35}^A$ & $0$ & $1$ & $1$\\
\bottomrule
\end{tabular*}
\captionof{table}{The values of the charges $Q_1$, $Q_2$ and $R_0$ for the fields of PdP$_{3b}^{\Omega}$ and PdP$_{3c}^{\Omega_2}$, which are the same. $Q_1$ and $Q_2$ refer to the charges under the flavour symmetry $U(1)_1 \times U(1)_2$, while $R_0$ is an allowed non-superconformal choice of the $R$-charge.}\label{tab}
\end{center}

\section{Discussion}

We have shown that the value of the $a$-charge of the superconformal fixed point of the unoriented PdP$_{3c}^{\Omega_2}$ is smaller than expected and this gives strong evidence that there is an RG flow from the UV to the IR. On the fixed point in the IR, the $R$-charges and $a$-charge are exactly those of the unoriented PdP$_{3b}^{\Omega}$. Moreover, the two theories share the same field content and have identical charges under the same global symmetry for any finite $N$. As a consequence, 't Hooft anomalies match in the IR as well as the superconformal index and thus we conjecture that in the IR the two unoriented theories describe the same physics.

A natural question that one can ask is whether the PdP$_{3b}^{\Omega}$ and PdP$_{3c}^{\Omega_2}$ theories are connected by an exactly marginal deformation. A hint in this direction comes from the fact that the two theories differ only because of superpotential terms. This implies that if one turns on in the PdP$_{3c}^{\Omega_2}$  theory  a deformation $\alpha ( W_{3b}^{\Omega} - W_{3c}^{\Omega_2})$, the resulting theory must have a superconformal fixed point for any value of $\alpha$, with the same value of the $R$-charges. While for  $\alpha=0$ and $\alpha=1$ the theory results from an orientifold projection, it would be very interesting to investigate the origin of the other superconformal theories. The existence of exactly marginal deformations and the structure of the conformal manifold can be deduced from the superconformal index, whose computation we leave as an open project.\footnote[1]{We thank the referee for pointing out this possibility to us.}

As far as the gravity side of the AdS/CFT correspondence is concerned, the mechanisms known in the literature to produce RG flows in the context of holographic field theories do not seem to explain our result. In particular, the flow described above is not due to mass deformations~\cite{Bianchi:2014qma, Bianchi:2020fuk}, to a Higgs mechanism~\cite{Feng_2003} or to the introduction of fractional branes~\cite{Gubser:1998fp}. We can expect some kind of kink solution interpolating between two asymptotic geometries like in~\cite{Freedman:1999gp}, but the metrics of the CY complex cones over PdP$_{3b}$ and PdP$_{3c}$ are unknown and thus we do not have control of the bulk theory. Progress on the gravity side of the orientifold theories discussed in this letter would also allow  one to investigate holographically the marginal deformation discussed above.

Since orientifolds realized in the dimer by fixed points do not break the $\textrm{U}(1)^2 \times \textrm{U}(1)_R$ mesonic symmetry~\cite{Franco_2006}, the $\textrm{U}(1)$ broken by the $\Omega_2$ orientifold must be baryonic. The picture is thus that the configuration of branes and orientifold planes in the PdP$_{3c}^{\Omega_2}$ theory breaks, together with a baryonic $\textrm{U}(1)$, also conformal symmetry in the UV, but makes the theory flow to a different IR fixed point that is the one of the  PdP$_{3b}^{\Omega}$ theory, with which PdP$_{3c}^{\Omega_2}$ shares all the symmetries. In the bulk, we can thus expect form fluxes that make the volume increase so that only asymptotically the metric is AdS. The scale associated to the flow can be identified with the size of the cycles wrapped by the branes that generate the fluxes. Solutions with broken baryonic $\textrm{U}(1)$'s associated to branes wrapping cycles have been discussed in the literature~\cite{Butti:2004pk, Maldacena:2009mw}.

Another direction that can be explored is the possibility that the duality is a consequence of the PdP$_{3b}$ and PdP$_{3c}$ parent theories being connected by specular duality~\cite{Hanany:2012hi,Hanany_2012}, which in general is a map between theories with the same master space. In the case of  theories whose toric diagram has only one internal point, like the ones we are discussing, specular duality exchanges mesonic and anomalous baryonic symmetries. The chain of maps that relate the two parent theories is more precisely a specular duality followed by a Seiberg duality~\cite{Seiberg:1994pq} and again another specular duality. One could even investigate the possibility that a chain of Seiberg dualities relates the two unoriented theories. Seiberg dualities in the case in which  (anti-)symmetric fields are present have been considered in~\cite{Pouliot:1995me, Berkooz:1995km}, where one describes them as mesons of a new confining symplectic or orthogonal gauge group. The problem of this approach is that one needs to add a gauge group going towards the UV and that can not describe the flow from PdP$_{3b}^{\Omega}$ to PdP$_{3c}^{\Omega_2}$.

We expect to find other examples of pairs of orientifolds sharing the same features of the theories discussed in this letter. This would allow us to understand the physical origin of this infrared duality. However, as the number of gauge groups increases, it becomes computationally harder to find the exact local maximum of the $a$-charge. To give more evidence that the unoriented PdP$_{3c}^{\Omega_2}$ flows to PdP$_{3b}^{\Omega}$ in the IR, we plan to study the moduli spaces of the two unoriented theories. Another possible line of investigation would be to check whether S-duality and strong coupling effects are involved, along the lines of~\cite{GarciaEtxebarria:2012qx, Garcia-Etxebarria:2013tba, Garcia-Etxebarria:2015hua, Garcia-Etxebarria:2016bpb}.

\section*{Acknowledgments}
We thank M. Bertolini, D. Bufalini and S. Meynet for useful comments and in particular M. Bianchi for discussions and suggestions at various stages of the project and for a careful reading of the manuscript.

\bibliographystyle{elsarticle-num} 
\bibliography{PdPFlow_PLB}

\end{document}